\documentclass[aps,twocolumn,preprintnumbers,showpacs,showkeys]{revtex4}
\usepackage{epsfig}
\usepackage{amssymb,amsmath,amsfonts,amsthm,graphicx,psfrag}
\graphicspath{{./Figures/}}                 
\newcommand{\noi}{\noindent}
\newcommand{\beq}{\begin{equation}}
\newcommand{\eeq}{\end{equation}}
\newcommand{\bea}{\begin{eqnarray}}
\newcommand{\eea}{\end{eqnarray}}

\newcommand{\Sec}[1]{Section~\ref{#1}}
\newcommand{\Eq}[1]{Eq.~(\ref{#1})}
\newcommand{\caa}{{\cal A}}

\newcommand{\vp}{{\vec p}}

\newcommand{\re}{\operatorname{\mathfrak{Re}}}
\newcommand{\tr}{\operatorname{Tr}}

\newcommand{\bc}{{\it bc~}}

\newcommand{\aleq}{\mbox{}_{\textstyle \sim}^{\textstyle < }}
\newcommand{\ageq}{\mbox{}_{\textstyle \sim}^{\textstyle > }}

\newcommand{\hatm}{\widehat{m}}
\begin{document}

\preprint{ITEP-LAT/2011-01}

\title{Lattice QCD gluon propagators near transition temperature
}

\author{V.~G.~Bornyakov}
\affiliation{High Energy Physics Institute, 142 281 Protvino, Russia \\
and Institute of Theoretical and Experimental Physics, 117259 Moscow, Russia}

\author{V.~K.~Mitrjushkin}
\affiliation{Joint Institute for Nuclear Research, 141980 Dubna, Russia \\
and Institute of Theoretical and Experimental Physics, 117259 Moscow, Russia}

\date{01.03.2011}

\begin{abstract}

Landau gauge gluon propagators are studied numerically in the $SU(3)$
gluodynamics as well as in the full QCD with the number of flavors
$n_F=2$ using efficient gauge fixing technique. We compare these
propagators at temperatures very close to the transition point in two
phases : confinement and deconfinement. The electric mass $m_E$ has
been determined from the momentum space longitudinal gluon  propagator.
Gribov copy effects are found to be rather strong in the gluodynamics,
while in the full QCD case they are weak ("Gribov noise"). Also we
analyse finite volume dependence of the transverse and longitudinal
propagators.

\end{abstract}

\keywords{Lattice gauge theory, gluon propagator, ghost propagator,%
Gribov problem, simulated annealing}

\pacs{11.15.Ha, 12.38.Gc, 12.38.Aw}

\maketitle

\section{Introduction}
\label{sec:introduction}

The transition from the confinement to the deconfinement phase is one of
the most interesting features of QCD at finite temperature. This transition
separates a low-temperature confinement phase from a high-temperature --
quark-gluon plasma  -- phase, where color charges should be deconfined.
The existence of this transition has been confirmed by recent observations
of the collective effects in ultrarelativistic heavy-ion collisions
(see, e.g., the review \cite{Gyulassy:2004zy}).

The non-perturbative study of gauge variant propagators is of
interest for various reasons.  These propagators are expected to show
different behavior in each phase and, therefore, to serve as a useful
'order parameters', detecting the phase transition point $T_c$.
One expects that their study can shed the light on the mechanism
of the confinement-deconfinement transition \cite{Kugo:1979gm,
Gribov:1977wm, Zwanziger:1991gz, Zwanziger:1993dh}.  Another reason
is that for the reliable phenomenological analysis of high-energy
heavy-ion collision data, it is important to obtain information on the
momentum dependence of the gluon propagators, especially in the infrared
region (see, e.g., \cite{Baier:1996kr,Baier:1996sk,Gyulassy:2003mc,
Kovner:2003zj,Wang:2000uj}).  Finally, the non-perturbatively calculated
lattice propagators are to be used to check the correctness of various
analytical methods in QCD, for example, Dyson--Schwinger equations method
at finite temperature \cite{Maas:2005hs,Maas:2005ym,Cucchieri:2007ta}.

Last years a number of papers have been dedicated to the lattice study
of the finite temperature $SU(2)$ pure gauge gluon propagators in
Landau gauge (see, e.g., papers \cite{Heller:1995qc, Heller:1997nqa,
Cucchieri:2001tw, Cucchieri:2007md, Cucchieri:2007ta, Fischer:2010fx,
Bornyakov:2010nc,Cucchieri:2011ga}). However, the finite temperature
$SU(3)$ gluon propagators - especially in the presence of dynamical
fermions - are much less studied (see papers \cite{Mandula:1987cp,
Nakamura:2002ki,
Nakamura:2003pu, Fischer:2010fx} for the pure gauge theory and
\cite{Furui:2006py} for the full QCD). 

In this paper we study the transverse (magnetic) and longitudinal
(electric) gluon propagators both in the pure gauge $SU(3)$ theory and
in the full QCD with the number of flavors $n_F=2$. The main goal of our
work is to compare the momentum behavior of these propagators very close
to the transition point, slightly above $T_c$ (deconfinement phase) and
slightly below $T_c$ (confinement phase).  We compare our results for the
gluon propagators computed in gluodynamics and in full QCD at coinsiding
values of the ratio $T/T_c$ where $~T_c~$ is a corresponding transition
temperature.  For both theories we have chosen temperatures $T$ very close
to the corresponding transition point : $~T/T_c=0.97~$ and $~T/T_c=1.02~$.
Of special interest in this study are the infrared mass scale parameters
('screening masses'): $m_E$ ('electric') and $m_M$ ('magnetic').

We apply effective gauge fixing algorithm - simulated annealing - and
generate few gauge copies for every configuration to reduce Gribov
copy effects. For  gluodynamics we apply additionally flips between
sectors of the Polyakov loops \cite{Bogolubsky:2005wf}. The flips were
applied to this theory for the first time. Also we  check finite volume
effects comparing results for two lattice sizes.

\vspace{2mm}

\Sec{sec:definitions} contains main definitions as well as some details of
simulations and gauge fixing procedure we use.  Volume and temperature
dependence of the propagators for both pure gauge theory and for QCD
are discussed in \Sec{sec:prop_mom}.  \Sec{sec:screening} is dedicated
to the discussion of the screening masses and \Sec{sec:conclusions}
is reserved for conclusions and discussion.

\hfill

\section{Gluon propagators: the definitions and simulation details}
\label{sec:definitions}

For the study of the gluon propagators in the pure gauge theory we
employ the standard Wilson action $S_W$ with $\beta = 6/g_0^2 \,$ where
$~g_0$ is a bare coupling constant.  To define the spacing $a$ as a
function of $\beta$ in the pure gauge case we have used Necco--Sommer
parametrization \cite{Necco:2001gh} with the popular choice of the Sommer
scale $r_0=0.5$fm.  In what follows we will refer to this theory as 
$n_F=0$ theory.

To study the effect of the quarks on the gluon propagator we computed
propagators on configurations generated with the gauge action $S_W$
and  $n_F=2$ dynamical flavors of nonperturbatively $O(a)$ improved
Wilson fermions (clover fermions).  The configurations were produced
by the DIK collaboration \cite{Bornyakov:2009qh}  with BQCD code
\cite{Nakamura:2010qh}.  The dimensionless quantities $r_0/a$ and 
$ m_\pi r_0$ are taken from results of QCDSF collaboration.
To convert them into physical units we use the value $r_0=0.467(15)$
fm obtained by QCDSF \cite{Gockeler:2005rv}.
The improvement coefficient $c_{SW}$ was determined nonperturbatively 
\cite{Jansen:1998mx}.  In the
rest of the paper this theory will be refered to as $n_F=2$ theory.

\vspace{2mm}

Our calculations were performed on the asymmetric lattices with lattice
volume $V=L_4\cdot L_s^3$, where $L_4$ is the number of sites in the
$4th$ direction. The temperature $T$ is given by $~T=1/aL_4~$ where $a$
is the lattice spacing.  We employ the standard definition of the lattice
gauge vector potential $\caa_{x\mu}$ \cite{Mandula:1987rh} :

\beq
\caa_{x\mu} = \frac{1}{2i}\Bigl( U_{x\mu}-U_{x\mu}^{\dagger}
\Big)_{\rm traceless} \equiv A_{x\mu}^a T^a ~,
\label{eq:a_field}
\eeq

\noi where link variables  $U_{x\mu}\in SU(3)$ transform under
gauge transformations $g_x \in SU(3)$ as follows : $ U_{x\mu}
\stackrel{g}{\mapsto} U_{x\mu}^{g} = g_x^{\dagger} U_{x\mu} g_{x+\mu}$.

The Landau gauge condition is given by
$~(\partial \caa)_{x} = \sum_{\mu=1}^4 \left( \caa_{x\mu}
- \caa_{x-\hat{\mu};\mu} \right)  = 0 ~$
which is equivalent to finding an extremum of the gauge functional

\beq
F_U(g) = ~\frac{1}{4V}\sum_{x\mu}~\frac{1}{3}~\re\tr~U^{g}_{x\mu} \;,
\label{eq:gaugefunctional}
\eeq

\noi  with respect to gauge transformations $g_x~$.

\vspace{2mm}

The bare gluon propagator $D_{\mu\nu}^{ab}(p)$ is given  by

\beq
D_{\mu\nu}^{ab}(p) = \frac{a^2}{g_0^2}
\Big\langle \widetilde{A}_{\mu}^a(k) \widetilde{A}_{\nu}^b(-k) \Big\rangle~,
\label{eq:gluonpropagator}
\eeq

\noi where $\widetilde{A}(k)$ represents the Fourier transform of the
gauge potentials according to \Eq{eq:a_field} after having fixed the
gauge.  The physical momenta $p$ are given by $~p_{i}=(2/a) \sin{(\pi
k_i/L)}, ~~p_{4}=(2/a) \sin{(\pi k_4/L_4)}, ~~k_i \in (-L/2,L/2], k_4 \in
(-L_4/2,L_4/2]$.

In what follows we consider only {\it soft} modes $p_4=0$.  The hard
modes ($p_4 \ne 0$) have an {\it effective} thermal mass $2\pi Tn$
and behave like massive particles.

On the asymmetric lattice there are two tensor structures for the gluon
propagator \cite{Kapusta}~:

\beq
D_{\mu\nu}^{ab}(p)=\delta_{ab} \left( P^T_{\mu\nu}(p)D_{T}(p) +
P^L_{\mu\nu}(p)D_{L}(p)\right)\,,
\eeq

\noi where (symmetric) orthogonal projectors $P^{T;L}_{\mu\nu}(p)$
are defined at $p=(\vec{p}\ne 0;~p_4=0)$ as follows

\bea
P^T_{ij}(p)&=&\left(\delta_{ij} - \frac{p_i p_j}{\vec{p}^2} \right),\,
~~~P^T_{\mu 4}(p)=0~;
\\
P^L_{44}(p) &=& 1~;~~P^L_{\mu i}(p) = 0 \,.
\eea

\noi Therefore, two scalar propagators - longitudinal $D^{L}(p)$ and
transverse $D^T(p)$ -  are given by

\beq
D^T(p) = \frac{1}{16} \sum_{a=1}^{8}\sum_{i=1}^{3}D_{ii}^{aa}(p);
~~D^L(p) = \frac{1}{8}\sum_{a=1}^{8} D_{44}^{aa}(p) .
\eeq

\noi For $\vec{p} = 0$ propagators $D^{T}(0)$ and $D^L(0)$ are given by

\beq
D^T(0) = \frac{1}{24} \sum_{a=1}^8 \sum_{i=1}^{3} D^{aa}_{ii}(0) ;
~~D^L(0) = \frac{1}{8} \sum_{a=1}^8 D^{aa}_{00}(0)  .
\eeq

\noi The transverse propagator $D^T(p)$  is associated  to magnetic
sector, and the longitudinal one $D^L(p)$ - to electric sector.

\vspace{2mm}

In the case of the $n_F=0$ theory we employ for gauge fixing the $Z(3)$
flip operation as has been proposed in \cite{Bogolubsky:2005wf}.
It consists in flipping all link variables $U_{x\mu}$ attached
and orthogonal to a 3d plane by multiplying them with $\exp{\{\pm
2\pi i/3\}}$.  Such global flips are equivalent to non-periodic gauge
transformations and represent an exact symmetry of the pure gauge
action $S_W$.  At finite temperature we apply flips only to directions
$\mu=1,2,3$.  As for the $4$th direction,
we stick to the sector with $ |arg~P| < \pi/3$ which provides maximal
values of the functional $F$. Therefore, the flip operations combine
for each lattice field configuration the $3^3$ distinct gauge orbits
of strictly periodic gauge transformations into one larger gauge orbit.

All details of our gauge fixing procedure - FSA algorithm - are described
in our recent papers \cite{Bogolubsky:2007bw, Bornyakov:2008yx,
Bornyakov:2009ug, Bornyakov:2010nc}. The  features of the
simulated annealing algorithm application specific for $SU(3)$
group are essentially same as in  \cite{Bogolubsky:2007ud} where
this algorithm was applied to $SU(3)$ theory for the  first time.
For every configuration we produce two gauge copies per  flip-sector;
therefore we have in total $N_{copy}=54$ copies. We take the copy with
maximal value of the functional (\ref{eq:gaugefunctional}) as our best
estimator of the global maximum and denote it as best (``\bc'') copy.

\vspace{1mm}

In the case of the $n_F=2$ theory the global $Z(3)$ transformations do
not anymore make part of the symmetry group of the action (only one
flip-sector remains). In this case we have made $N_{copy}=10$ gauge copies.

\vspace{1mm}

To suppress 'geometrical' lattice artifacts, we have applied the
``$\alpha$-cut'' \cite{Nakagawa:2009zf}, i.e.  $~\pi k_i/L_s < \alpha~$,
for every component, in order to keep close to a linear behavior
of the lattice momenta $p_i \approx (2 \pi k_i)/(aL_s), ~~k_i \in
(-L_s/2,L_s/2]$.  We have chosen $\alpha=0.5$. Obviously, this cut
influences large momenta only.
We did not employ the cylinder cut in this work.

\vspace{1mm}

The values of $T/T_c$, $\beta$, $\kappa$ and lattice sizes are
given in Table \ref{tab:statistics1}.
Number of independent configurations in all cases equals $\sim 200$.


\begin{table}[ht]
\begin{center}
\begin{tabular}{|c|c||c||c|c|} \hline
\multicolumn{2}{|c||}{} & \multicolumn{1}{|c||}{$n_F=0$} &
\multicolumn{2}{|c|}{$n_F=2$}
\\ \hline 
 $~T/T_c~$ & $L_4\cdot L_s^3$ & $\beta$ & $\beta$ & $\kappa$ \\
\hline
 0.97  & $8\cdot 16^3$ & 6.044 & 5.25 & 0.1341  \\
 1.02  & $8\cdot 16^3$ & 6.075 & 5.25 & 0.1345  \\
\hline
 0.97  & $8\cdot 24^3$ & 6.044 & 5.25 & 0.1341  \\
 1.02  & $8\cdot 24^3$ & 6.075 & 5.25 & 0.1345  \\
\hline
\end{tabular}
\end{center}
\caption{Run parameters.
}
\label{tab:statistics1}
\medskip \noindent
\end{table}


For $n_F=0$ theory the critical value $\beta_c$ (corresponding to the
$1st$ order phase transition) has been taken from \cite{Boyd:1996bx}.

In the case of $n_F=2$ theory the simulations were made
\cite{Bornyakov:2009qh} at fixed $\beta=5.25$ and varying $\kappa$. The
pseudocritical values $\kappa_c$ (and thus the transition temperatures
$T_c$) of the deconfining and chiral symmetry restoration transitions
have been determined \cite{Bornyakov:2009qh} by the maximum of the
Polyakov loop and chiral condensate susceptibilities, respectively. It
has been found \cite{Bornyakov:2009qh}  that these two maxima coinside
within numerical precision at $\kappa_c = 0.1343$.

The pion is heavy at values of $\kappa$ we have chosen ($m_{\pi}
\sim 1$ Gev). However, it is known that the effect of the quarks is
rather strong even at these values of $m_\pi$. The transition is
a crossover, rather than a $1st$ order phase transition.
Furthermore, the transition temperature is shifted
to substantially lower values in comparison to the gluodynamics
\cite{Karsch:2000kv,Ejiri:2009hq, Bornyakov:2009qh} and the string
breaking phenomenon is observed at $T<T_c$ \cite{Detar:1998qa,
Bornyakov:2002iv}.

\section{Gluon propagators : numerical results}
\label{sec:prop_mom}

We define the renormalized propagators $D^{T}_{ren}(p)$ and $D^L_{ren}(p)$
in such a way that their dressing functions are equal to unity at the
normalization point $\mu=2.5$ Gev. In the rest of this paper we will omit
the subscript '{\it ren}'; therefore, $D_T(p)$ and $D_L(p)$ will denote
transverse and longitudinal renormalized propagators, respectively.

\vspace{2mm}


\begin{table}[ht]
\begin{center}
\begin{tabular}{|c|c||c|c|c|c|} \hline
\multicolumn{2}{|c||}{} & \multicolumn{2}{|c|}{$n_F=0$} &
\multicolumn{2}{|c|}{$n_F=2$}
\\ \hline 
 $~T/T_c~$ & $L_4\cdot L_s^3$ & $Z_M$ & $Z_E$ & $Z_M$ & $Z_E$ \\
\hline
 0.97  & $8\cdot 16^3$ & 1.03(1) & 1.00(2) & 1.12(2) & 1.14(2)    \\
 1.02  & $8\cdot 16^3$ & 1.01(1) & 1.00(1) & 1.13(2) & 1.16(2)    \\
\hline
 0.97  & $8\cdot 24^3$ & 1.04(1) & 1.00(1) & 1.14(1)  & 1.14(1)  \\
 1.02  & $8\cdot 24^3$ & 1.04(1) & 1.03(1) & 1.15(1)  & 1.17(1)  \\
\hline
\end{tabular}
\end{center}
\caption{Renormalization constants $Z_{M/E}$.
}
\label{tab:renorm}
\medskip \noindent
\end{table}


To compute the renormalization constants $Z_M$ and $Z_E$ we fitted the
propagators at momenta $p$ in the range between 2.3 and 2.7 GeV to  a
polinomial function. They are shown in Table \ref{tab:renorm} for  both
$n_F=0$ and $n_F=2$ theories. 
As expected, there is no volume dependence of these constants.
In the case of the pure gauge theory all $Z_M$ and $Z_E$ are very
close to unity, which is in agreement with the results for the
bare gluon propagators at zero temperature in pure gauge $SU(3)$
theory \cite{Sternbeck:2005tk}. In the $n_F=2$ case all
renormalization constants are substantially larger. This deviation
($\sim 15\%$) might be explained, at least, partially,  by the quark
contribution. Another possible explanation might be the larger spacing
(or smaller $\beta$) effect. 

For both theories every constant, i.e., $Z_M$ and $Z_E$, is practically
the same above and below $T_c$. This means that ultraviolet parts of
$D_T(p)$ and $D_L(p)$ are not 'phase-sensitive'.

\subsection{On Gribov copy effects}

In our study of the Gribov copy effects we follow the approach described
briefly in Section \ref{sec:definitions} and - in more details - in
our papers \cite{Bakeev:2003rr,Bogolubsky:2005wf, Bogolubsky:2007bw,
Bornyakov:2008yx, Bornyakov:2009ug, Bornyakov:2010nc} dedicated to the
Gribov copy effects in SU(2) gluodynamics at zero and nozero temperatures.

\vspace{1mm}

We find that in the case of the $n_F=0$ theory, similar to $SU(2)$
gluodynamics, the Gribov copy influence still remains a serious problem
in the lattice calculations, at least, at small momenta.
These effects are very strong for $D_T(p)$ in the infrared and weak
for $D_L$(p) for all momenta. 

It is worthwhile to note that the effect of different flip-sectors plays
a very important role in this case, even more important than additional
gauge copies in each sector.  The details of this study will be published
elsewhere.

In contrast, in the case of dynamical fermions ($n_F=2$) the Gribov copy
effects look completely negligible as compared to the pure gauge case.
The propagators do not change within errorbars ("Gribov noise").
However, it is worthwhile to note that our lattice sizes are comparatively
small. One cannot exclude that these effects might become stronger
for larger lattice sizes and/or smaller lattice spacings.

\subsection{Volume dependence}

\begin{figure}[tb]
\centering
\includegraphics[width=6.8cm,angle=270]{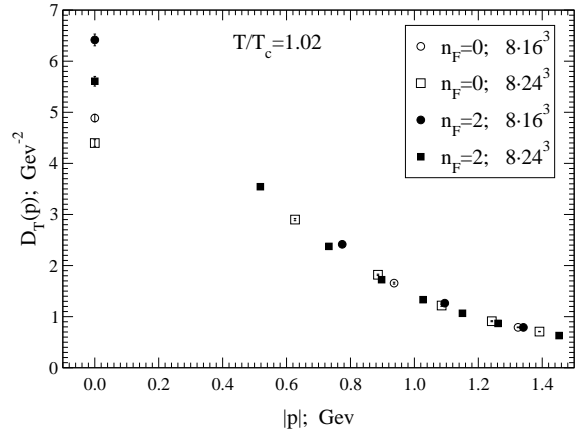}
\caption{The momentum dependence of the transverse
propagator $D_T(p)$ at $T/T_c=1.02$ for $|p| \le 1.5$ Gev
for two lattice sizes: $L_s=16$ and 24.
}
\label{fig:glpm_nF2_nF0_2Vol_1p02_IR}
\end{figure}

We find that for the transverse propagators $D_T(p)$ the finite volume 
effects are rather strong at zero momentum but they become very weak 
with increasing $|p|$.  As an illustration, in Figure
\ref{fig:glpm_nF2_nF0_2Vol_1p02_IR} we show the momentum dependence of
the $n_F=0$ and $n_F=2$ transverse propagators at comparatively small
momenta ($|p| < 1.5$ GeV) at $T/T_c=1.02$ on the lattices $8\cdot 16^3$
and $8\cdot 24^3$.  No volume dependence is observed also for $|p|>
1.5$ Gev. The similar situation takes place  for 
$T/T_c=0.97$.

\begin{figure}[tb]
\centering
\includegraphics[width=6.8cm,angle=270]{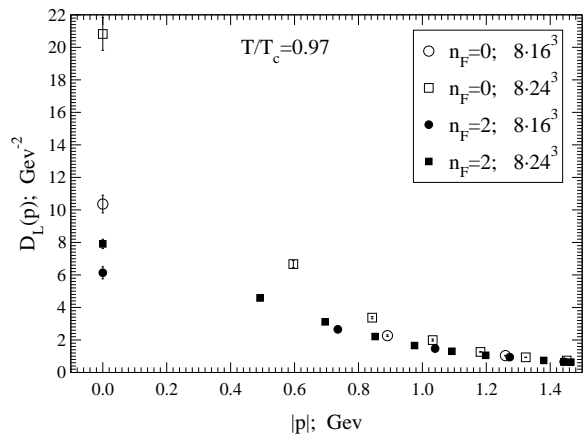}
\caption{The momentum dependence of the longitudinal
propagator $D_L(p)$ at $T/T_c=0.97$ for $|p| \le 1.5$ Gev.
}
\label{fig:glpe_nF2_nF0_2Vol_0p97_IR}
\end{figure}

The volume dependence of the longitudinal propagators $D_L(p)$ is
different. One can see from Figure~\ref{fig:glpe_nF2_nF0_2Vol_0p97_IR}
that the sign of the finite volume effects differs from that for
$D_T(p)$ case : while the transverse propagator decreases with increasing
volume, the longitudinal one is increasing. This is in agreement with
$SU(2)$ theory  \cite{Bornyakov:2010nc}. Another observation is that
the finite volume effects are much more substantial for $n_F=0$ theory 
than for $n_F=2$ one. At $p=0$ the quenched theory propagator increases 
by factor two when $L_s$ increases from $16$ to $24$ as can be seen 
from Figure~\ref{fig:glpe_nF2_nF0_2Vol_0p97_IR} at $T/T_c=0.97$. 
This increase is similar to that observed in $SU(2)$ theory
\cite{Bornyakov:2010nc,Cucchieri:2007md}. The finite volume effects
decrease fast with increasing $|p|$ both for $n_F=0$ and for $n_F=2$ 
theories.  At $T/T_c=1.02$ the corresponding figure looks in a similar way.
    
In what follows we show data for $L_s=24$ lattices only.

\vspace{2mm}

\subsection{Temperature dependence}

\begin{figure}[tb]
\centering
\includegraphics[width=6.8cm,angle=270]{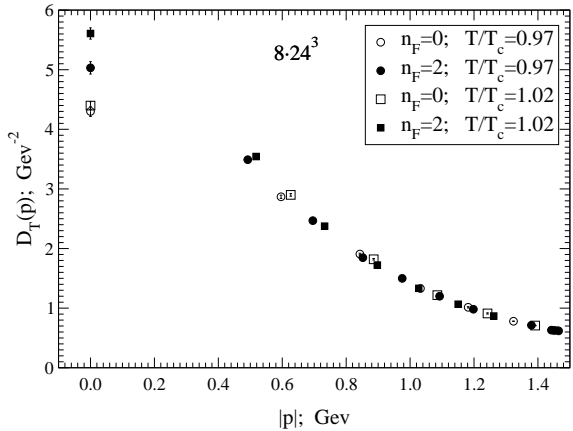}
\caption{The momentum dependence of the transverse
propagator $D_T(p)$ at two temperatures for $|p| \le 1.5$ Gev.
}
\label{fig:glpm_nF2_nF0_24x08_2temp_IR}
\end{figure}

We find that the transverse propagators $D_T(p)$ are not very much 
sensitive to the crossing of the transition point $T_c$, at least, at 
$p\ne 0$. 
In Figure \ref{fig:glpm_nF2_nF0_24x08_2temp_IR} where we present results 
for $D_T(p)$ at  $T/T_c=0.97$ and $T/T_c=1.02$ for both $n_f=0$ and
$n_f=2$ theories.  We observe that in the case of 
$n_f=0$ theory the temperature dependence is very weak at all momenta. 
(Similar conclusion was made in \cite{Fischer:2010fx}.)
The same is true  for $n_f=2$ theory at nonzero momenta. But, contrary to
$n_f=0$ case,  for $p=0$ substantial change in the propagator is observed 
as can be seen from Figure \ref{fig:glpm_nF2_nF0_24x08_2temp_IR}. 
This is one of the clear effects of dynamical quarks.

Another interesting observation is that one can {\it not} see any
substantial influence of the dynamical fermions on the momentum dependence
of the transverse propagators for $|p| \ne 0$. Moreover, for the momenta
$|p| \, \ageq \, 0.8 $ Gev all four propagators coinside within
errorbars.  
It would be interesting to see how this situation will change
for {\it smaller} values of the pion mass $m_{\pi}$.
The values of propagators obtained at $p=0$ suggest that in $n_f=2$  theory
the transverse propagator is infrared enhanced in comparison with
the one in $n_f=0$  theory. This is in contrast with $T=0$ results 
\cite{Bowman:2004jm,Ilgenfritz:2006gp}.
But this conclusion should be checked in future studies on larger volumes.

\vspace{2mm}

\begin{figure}[tb]
\centering
\includegraphics[width=6.8cm,angle=270]{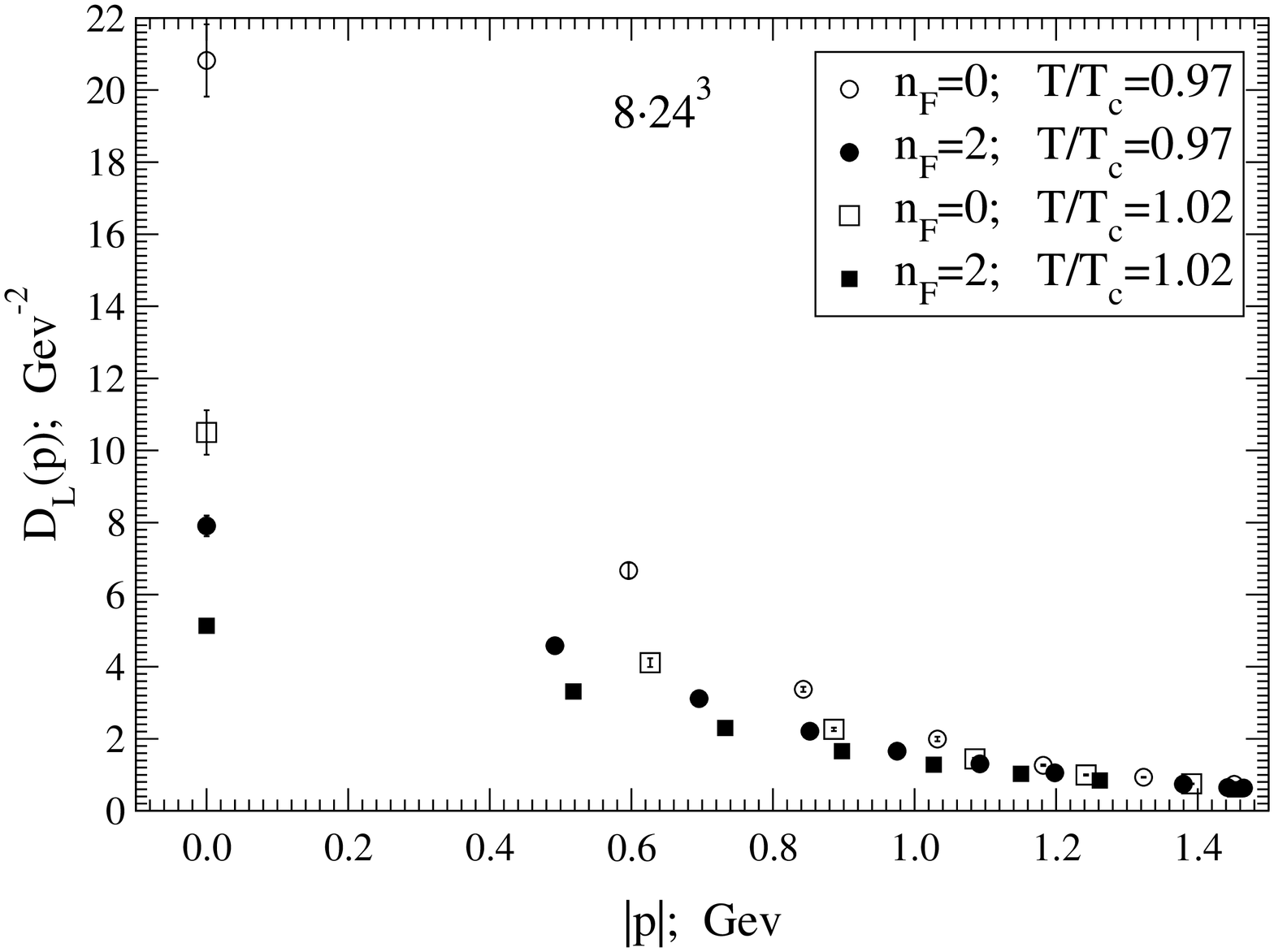}
\caption{The momentum dependence of the longitudinal
propagator $D_L(p)$ at two temperatures for $|p| \le 1.5$ Gev.
}
\label{fig:glpe_nF2_nF0_24x08_2temp_IR}
\end{figure}

In contrast, the longitudinal propagators $D_L(p)$ demonstrate much
more pronounced temperature dependence for  both theories, especially
in the infrared region  $|p| \le 1.5$ Gev. For both theories, i.e.,
$n_F=0$ and $n_F=2$, the propagators $D_L(p)$ differ significantly
also for nonzero values of momenta when temperature $T$ crosses the
transition point $T_c$ (see Figure \ref{fig:glpe_nF2_nF0_24x08_2temp_IR}).
Fast change of $D_L(p)$ across transition in $n_F=0$  theory was found recently
in \cite{Fischer:2010fx}.

\begin{figure}[tb]
\centering
\includegraphics[width=6.8cm,angle=270]{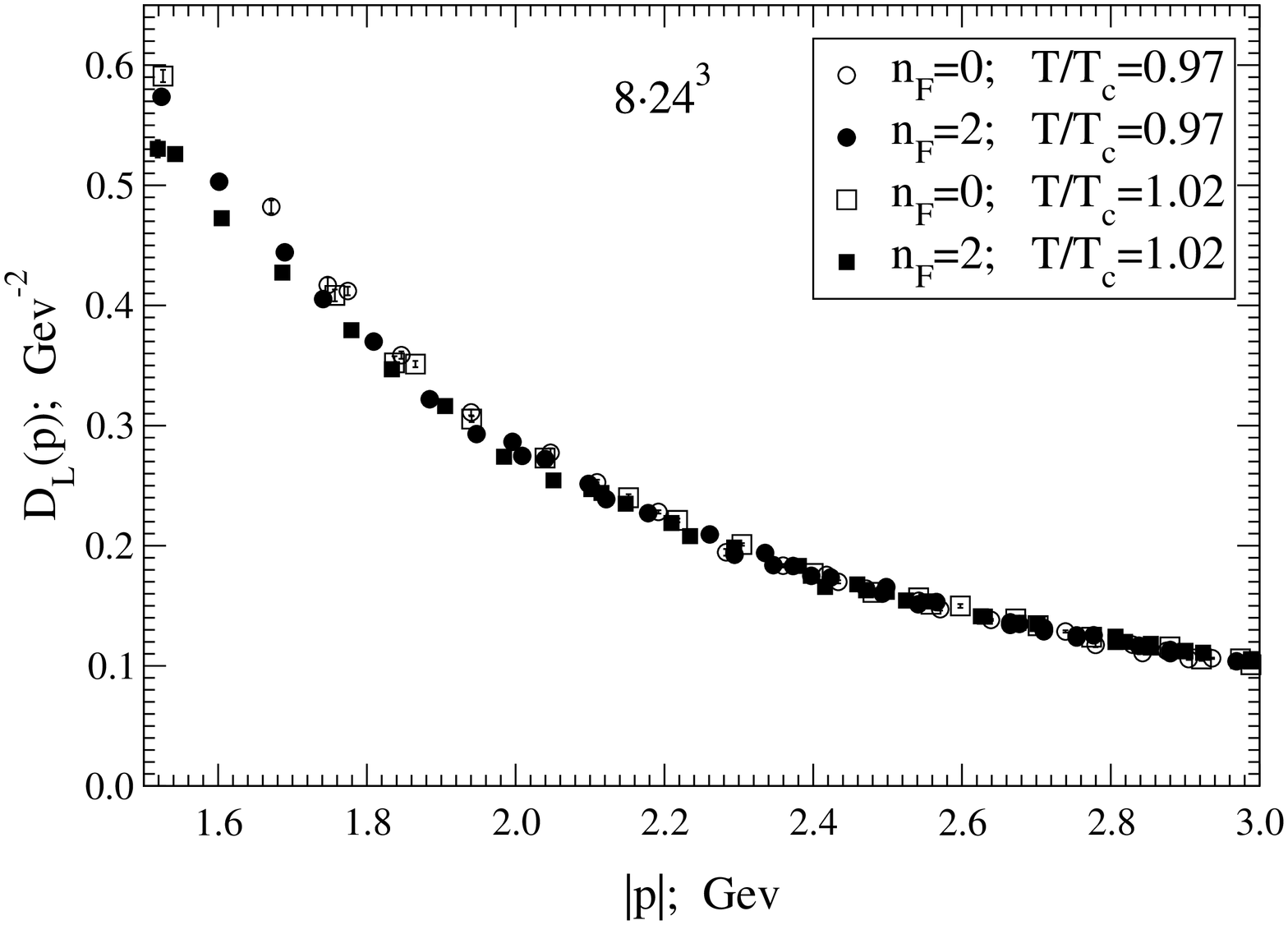}
\caption{The momentum dependence of the longitudinal
propagator $D_L(p)$ at two temperatures for $|p| > 1.5$ Gev.
}
\label{fig:glpe_nF2_nF0_24x08_2temp_UV}
\end{figure}

Fermions influence the longitudinal propagators up to values
of momenta $|p| \sim 1.8$ Gev what can be seen in Figure
\ref{fig:glpe_nF2_nF0_24x08_2temp_UV}. For larger values of $|p|$
we see no difference between $D_L(p)$ computed in $n_F=0$ and $n_F=2$ theories
as well as no temperature dependence.

\begin{figure}[tb]
\centering
\includegraphics[width=6.8cm,angle=270]{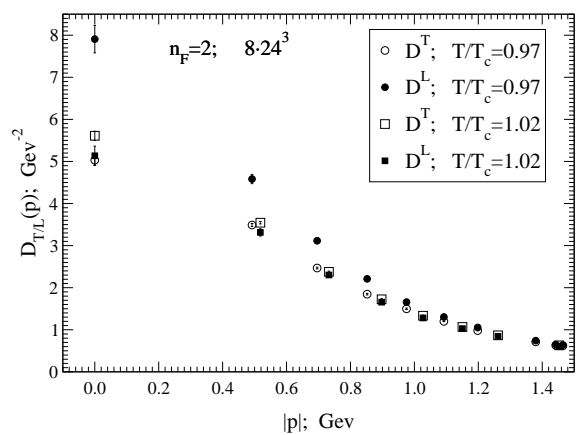}
\caption{The momentum dependence of $D_T(p)$  and $D_L(p)$
at two temperatures for $|p| < 1.5$ Gev.
}
\label{fig:glp_nF2_nF0_24x08_2temp_IR}
\end{figure}

Finally, we compare the momentum dependence of the
transverse and longitudinal propagators for both theories at different
temperatures. 
In Figure \ref{fig:glp_nF2_nF0_24x08_2temp_IR} 
the comparison of $D_T(p)$ and $D_L(p)$ in the
infrared region ($|p| < 1.5$ Gev) for $n_F=2$ theory is present.

Evidently, there is a large difference between $D_T(p)$ and $D_L(p)$
at small  momenta. However, this difference quickly decreases
with increasing $|p|$ and becomes zero (within errorbars) for $|p|
\ageq 1.2$ Gev.

\section{On the screening masses}
\label{sec:screening}

\begin{figure}[tb]
\centering
\includegraphics[width=6.8cm,angle=270]{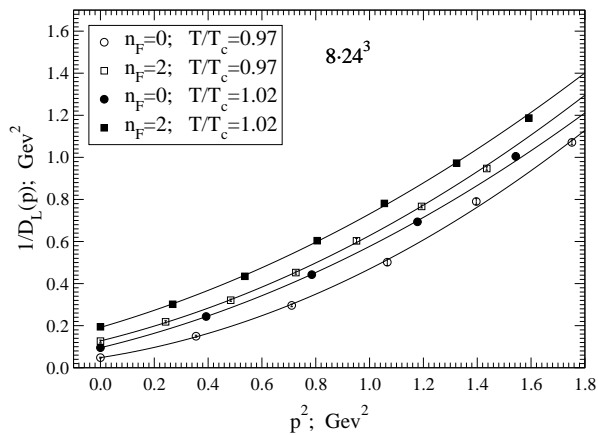}
\caption{The momentum dependence of the inverse propagators
$D_L^{-1}(p)$ for two different temperatures. Solid lines
correspond to fit to eq.(\ref{eq:massfit}).
}
\label{fig:glpe_inv_nF2_nF0}
\end{figure}

One of the interesting features of the finite temperature physics is the
appearance of the infrared mass scale parameters : $m_E$ ('electric')
and $m_M$ ('magnetic'). These parameters (or 'screening masses')
define screening of electric and magnetic fields at large distances
and, therefore, control the infrared behavior of $D_L(p)$ and $D_T(p)$.
The electric screening mass $m_E$ has been computed in the leading order
of perturbation theory long ago: $~m_E^2/T^2 = (1+n_F/6)g^2(T)~$ for
$SU(3)$ theory with $n_F$ flavors \cite{Nadkarni:1982kb, Nadkarni:1988fh}.
But at the next order the problem of the infrared divergencies has
been found.  On the other hand, the magnetic mass $m_M$ is entirely
nonperturbative in nature.  Thus a first-principles nonperturbative
calculations in lattice QCD should play an important role in the
determination of these quantities.

\vspace{1mm}

In the deep infrared region the momentum dependence of the longitudinal
propagator is expected to fit the pole-type behavior

\beq
D_L(p) \simeq \frac{C}{m_E^2 + p^2}~; \quad p\sim 0~.
\eeq

\noi Indeed, in the case of the pure gauge $SU(2)$ theory we
observed the linear behavior of the inverse longitudinal propagator
$D_L^{-1}(p)$ as a function of $p^2$ for small enough values of momenta
\cite{Bornyakov:2010nc}. Moreover, this linear dependence has been
observed both above and below $T_c$ and even at $T \simeq T_c$.

In the present work our lattice sizes $L_s$ do not permit us to reach
small enough values of $p^2$ where the linear $p^2$-behavior  of
$~D^{-1}_L(p)~$ could be found. Instead we have used for the fit
a somewhat symplified Stingl-like parametrization of the longitudinal
propagator \cite{Stingl:1985hx,Habel:1990tw}

\beq
D^{-1}_L(p) = C^{-1} \cdot (\widehat{m}_E^2 + p^2 + b |p|^4 )~.
\label{eq:massfit}
\eeq

\noi We emplyed this formula in the comparatively large momentum interval,
up to $p^2 \aleq 2$ Gev${}^2$, where $6$ data points were used 
for the fit for
every data set (see Figure \ref{fig:glpe_inv_nF2_nF0}). The parameter $b$
indicates the deviation from the simple pole-type behavior. We use the
parameter $\widehat{m}_E$ as an {\it estimator} of the electric screening
mass $m_E$.


\begin{table}[ht]
\begin{center}
\begin{tabular}{|c|c|c|c|c|} \hline
$T/T_c$ &  $n_F$   & $\widehat{m}_E/T$ & $C$     &  $bT^2$
\\ \hline\hline
0.97 & 0 &  1.725(126) & 5.03(49)  & 0.091(14) \\
0.97 & 2 &  2.09(15)   & 3.17(14)  & 0.054(6)  \\ \hline
1.02 & 0 &  1.851(77)  & 3.24(15)  & 0.050(5)  \\
1.02 & 2 &  2.24(9)    & 2.68(16)  & 0.045(7)  \\
\hline
\end{tabular}
\end{center}
\caption{Values of parameters  $\widehat{m}_E$, $C$ and $b$
obtained from fits to eq.(\ref{eq:massfit}).
}
\label{tab:electric_mass}
\medskip \noindent
\end{table}


The values of the fit parameters are presented in the Table
~\ref{tab:electric_mass}. In the case of the pure gauge theory we can
compare our value of $~\hatm_E/T~$ calculated at $~T/T_c=1.02~$ with
$~m_E/T~$ obtained in \cite{Nakamura:2003pu} for nearby temperature
$~T/T_c=1.05~$.  Both values are well consistent within errorbars.
Instead, in the unquenched QCD case our value of $~\hatm_E/T~$ at
$~T/T_c=1.02~$ differs by factor 1.5 approximately from $~m_E/T~$ 
calculated in $n_F=2$ QCD with KS fermions at the same value of $T/T_c~$
\cite{Furui:2006py}. The origin of this deviation is still to 
to be clarified.

As is well-known, one can hardly rely  on the perturbation theory at 
$T \sim T_c$. However, it is interesting to note that the ratio of 
two values $~m_E^2/T^2~$ (for $n_F=2$ and $n_F=0$) is consistent with 
the factor $(1+n_F/6)$.

\vspace{1mm}

Our main observation is that the dimensionless parameter $~\hatm_E/T~$
is practically {\it not} sensitive (within errorbars) to the crossing of
the transition point $T_c$. This statemant is true not only for $n_F=2$
theory where the crossover (or higher order phase transition)is expected
but also for the pure gauge theory where the existence of the $1st$
order phase transition is firmly established. Therefore, we conclude
that the electric mass can hardly be considered as an 'order parameter'
indicating the transition point $T_c$.

Note, that a similar situation have been observed in the case
of the finite temperature pure gauge $SU(2)$ theory where the values of
the parameter $~m_E/T~$ coinside for $~T/T_c=0.9~$ and $~T/T_c \simeq 1~$
\cite{Bornyakov:2010nc}.

\vspace{1mm}

In contrast, dimensionless
factors $C$ are much more sensitive to the crossing of the
transition point $T_c$.  This is true for  both $n_F=0$ and $n_F=2$
theories. The difference between $n_F=0$ and $n_F=2$ cases is more
quantitative than qualitative : the temperature variation of the parameter
$C$ in the pure gauge theory is much more strong as compared to the full
QCD case.

\vspace{1mm}

In fact, the only qunatity  which demonstrates  clear difference between 
$n_F=0$ and $n_F=2$ cases is the dimensionless parameter $~bT^2$.
Indeed, for the pure gauge theory the temperature jump of $bT^2$ is
rather strong, while for the $n_F=2$ case values of $bT^2$ above and below
$T_c$ coinside within errorbars.

\vspace{3mm}

The calculation of the magnetic screening mass $m_M$ looks a somewhat more
delicate problem. In the case of pure gauge $SU(2)$ theory it has been
shown that at $T>T_c$ the transverse propagator $D_T(p)$
has a maximum at $p \ne 0~$ \cite{Cucchieri:2001tw, Cucchieri:2007ta,
Fischer:2010fx}. 
This has been found also in $SU(3)$ gluodynamics \cite{Fischer:2010fx}.
Moreover, in our recent paper \cite{Bornyakov:2010nc} it
has been shown that $D_T(p)$ has a maximum not only at high temperatures
but even  at $T<T_c$.  The position of this maximum at $T$ close to
$T_c$ was found to be at about 400 Mev.  Therefore, the transverse
propagator $D_T(p)$ has a form which is {\it not} compatible with the
simple pole-type behavior, so for $m_M$ another, different from pole
mass, definition is necessary.  This definition has been proposed in
\cite{Bornyakov:2010nc}.

In paper \cite{Zahed:1999tg} it has been suggested that the proximity
of the Gribov horizon at finite temperature forces the transverse gluon
propagator $D_T(\vp,p_4=0)$ to vanish at zero three-momentum. If this
is the case, then the finite-temperature analog of Gribov formula
$~|\vp|^2/(|\vp|^4+M_M^4)~\equiv~1/(|\vp|^2+m^2_{e\! f\! f}(\vp))~$
suggests that the {\it effective} magnetic screening mass
$~m_{e\! f\! f}(\vp)~$ becomes infinite in the infrared (so called,
magnetic gluons 'confinement').

In the case of $SU(3)$ gauge group, both for $n_f=0$ and $n_f=2$
theories considered in this paper, we do not see a maximum of $D_T(p)$
at $p \ne 0$. The transverse propagators $D_T(p)$ look somewhat similar
to the longitudinal ones.  The most probable reason for this is that
our volumes are not large enough.  Note that also in the case of $SU(2)$
theory we did not observe the maximum at $p\ne 0$ on smaller lattices;
this maximum appeared only with increasing of lattice size $L_s$
\cite{Bornyakov:2010nc}.
To fit $D_T(p)$ we employed the same formula given in
eq.~(\ref{eq:massfit}) and same fitting range as for $D_L(p)$.  We found
that the fit is good when $p=0$ is excluded. Otherwise the $\chi^2/ndf$
increases up to $6$ and becomes significantly larger than in the case
of $D_L(p)$. We consider this as an indication of the maximum to be seen
when the volume is increased.

We conclude that for reliable definition of the magnetic screening mass
$m_M$ one needs larger values of the lattice size $L_s$.

\section{Conclusions}
\label{sec:conclusions}

In this work we studied numerically the behavior of the Landau gauge
transverse and longitudinal gluon propagators in the pure gauge
$SU(3)$ theory and in the theory with dynamical fermions with the number
of flavors $n_F=2$. We compare these two theories  in the close vicinity of
the transition temperature $T_c~$, slightly above and slightly below
corresponding transition temperature.  For  both theories
we have chosen temperatures $T$ in such a way that 
ratios $~T/T_c$ are equal to 0.97 and 1.02.
To our knowledge, this is the first study where two
theories are compared in confinement and deconfinement phases.

\vspace{1mm}

Let us summarize our findings.

\vspace{1mm}

In the case of the pure gauge theory the Gribov copy effects are rather
strong for $D_T(p)$ in the infrared and weak for $D_L$(p) for all momenta.
In contrast, in the case of dynamical fermions the Gribov copy effects 
look completely negligible (the so called "Gribov noise").  

\vspace{1mm}

The renormalization constants $Z_M$ and $Z_E$  are practically
the same above and below $T_c$ for  both theories.  It confirms 
that ultraviolet parts of $D_T(p)$ and $D_L(p)$ are not 'phase-sensitive'.

\vspace{1mm}

Both below and above $T_c$ the finite volume effects are rather strong for 
the transverse propagators $D_T(p)$  at zero momentum but they 
become weak fast with increasing $|p|$. This weakening is
slower for $n_f=2$ theory.  The volume dependence of the
longitudinal propagators $D_L(p)$ is 
different. First, the sign of the effect is different from that for  
$D_T(p)$ case.  Second, it is much more substantial for $n_F=0$  
theory than for $n_F=2$ one. At $p=0$ the quenched theory 
propagator increases by factor two with increasing volume as can 
be seen from Figure~\ref{fig:glpe_nF2_nF0_2Vol_0p97_IR}. 
This increase is similar to that observed in $SU(2)$ theory 
\cite{Bornyakov:2010nc,Cucchieri:2007md}. However, at $p\ne 0$ 
this dependence is not very much pronounced, both for $n_F=0$
and for $n_F=2$ theories.

\vspace{1mm}

We computed the electric mass $~m_E~$  defined in eq.(\ref{eq:massfit}).
It is practically {\it not} sensitive  to the
crossing of the transition point $T_c$. This is true not only for $n_F=2$
theory where the crossover (or higher order phase transition) is expected
but also for the $n_F=0$ case where the existence of the $1st$ order
phase transition is well established. Therefore, $m_E$ can hardly be
considered as an 'order parameter' indicating the transition point $T_c$.

In contrast, factors $C$ defined in eq.~(\ref{eq:massfit}) are much more
sensitive to the confinement-deconfinement transition for both $n_F=0$
and $n_F=2$ theories. However, the difference between $n_F=0$ and $n_F=2$
cases is more quantitative than qualitative : the temperature variation
of $C$ in the pure gauge theory is much more strong as compared to
QCD case. 
Note, that temperature dependence for $~m_E~$  and $C$  computed in  
$SU(2)$ gluodynamics \cite{Bornyakov:2010nc} was similar to those found 
here in $n_F=0$ theory. 

The only (dimensionless) parameter which demonstrates clear difference
between $n_F=0$ and $n_F=2$ theories near $T_c$ is $~bT^2$.  For the
pure gauge theory the confinement-deconfinement variation of $bT^2$
is rather strong. Instead, for the full QCD case values of $bT^2$ above
and below $T_c$ coinside within errorbars.

\vspace{1mm}

Calculation of the magnetic screening mass $m_M$ is  somewhat more 
delicate problem.  Our caution is based
on the experience of $m_M$ study in the $SU(2)$ theory.  Indeed, in the
case of the pure gauge $SU(2)$ theory the transverse propagator $D_T(p)$
has a local maximum at $p \ne 0$ in the deconfinement and even in the
confinement phases when lattice size $L_s$ is large enough and Gribov
copies are properly handled.  In our study of $SU(3)$ theories we do not
see a maximum of $D_T(p)$ at $p \ne 0$ and the transverse propagators
$D_T(p)$ look similar to the longitudinal ones. It is very
probable that this is because our lattice sizes are not large enough.
We conclude that for reliable definition of $m_M$ one needs larger values
of $L_s$ and $\beta$'s.

\subsection*{Acknowledgments}

This work was supported by the grant for scientific schools
NSh-6260.2010.2, by the Federal Special-Purpose Programme Cadres
of the Russian Ministry of Science and Education
and by the grant RFBR 09-02-00338-a.
The authors are thankful to M. M\"uller-Preussker for useful 
discussions.


\end{document}